\documentclass[iop,njp,reprint,superscriptaddress,showkeys,article-title]{revtex4-1} 
\usepackage[utf8]{inputenc}
\usepackage[UKenglish]{babel}

\usepackage[T1]{fontenc}
\usepackage{microtype}
\usepackage{amsmath,amssymb,mathtools,mleftright}
\usepackage{dsfont}  

\usepackage{graphicx,grffile,textcomp}
\graphicspath{{pics/}}
\newlength{\figwidth}

\usepackage{siunitx}
\usepackage{booktabs,tabularx,dcolumn}
\newcolumntype{d}[1]{D{.}{.}{#1}}

\usepackage{url,hyperref}
\usepackage[usenames,dvipsnames]{xcolor}
\hypersetup{colorlinks=true, linkcolor=BrickRed, urlcolor=blue!50!black, citecolor=blue!50!black}

\usepackage[capitalize]{cleveref}
\crefrangelabelformat{equation}{(#3#1#4)$-$(#5#2#6)}

\usepackage[normalem]{ulem}

\renewcommand\epsilon{\varepsilon}
\renewcommand\phi{\varphi}
\renewcommand\theta{\vartheta}
\renewcommand\rho{\varrho}

\renewcommand\geq{\geqslant}
\renewcommand\vec[1]{\textrm{\bfseries #1}}

\begin{document}

\title{Spontaneous trail formation in populations of auto-chemotactic walkers}

\author{Zahra Mokhtari}
\affiliation{Freie Universit{\"a}t Berlin, Department of Mathematics and Computer Science, \\ Arnimallee 6, 14195 Berlin, Germany}

\author{Robert I. A. Patterson}
\affiliation{WIAS, Mohrenstr. 39, 10117 Berlin, Germany}

\author{Felix H\"ofling}
\affiliation{Freie Universit{\"a}t Berlin, Department of Mathematics and Computer Science, \\ Arnimallee 6, 14195 Berlin, Germany}
\affiliation{Zuse Institute Berlin, Takustra{\ss}e 7, 14195 Berlin, Germany}

\date{\today}

\figwidth=3.4in 

\begin{abstract}
We study the formation of trails in populations of self-propelled agents that make oriented deposits of pheromones and also sense such deposits to which they then respond with gradual changes of their direction of motion.
Based on extensive off-lattice computer simulations aiming at the scale of insects, e.g., ants, we identify a number of emerging stationary patterns and obtain qualitatively the non-equilibrium state diagram of the model, spanned by the strength of the agent--pheromone interaction and the number density of the population.
In particular, we demonstrate the spontaneous formation of persistent, macroscopic trails,
and highlight some behaviour that is consistent with a dynamic phase transition.
This includes a characterisation of the mass of system-spanning trails as a potential order parameter.
We also propose a dynamic model for a few macroscopic observables, including the sub-population size of trail-following agents,
which captures the early phase of trail formation.
\end{abstract}

\maketitle

\section{Introduction}
Trail formation is known in populations of insects \cite{traniello1995trail, jeanne1981chemical, roces2009food, aron1993memory, jackson2004trail, czaczkes2015trail} and animal herds, as well as in colonies of microorganisms and, to some extent, also for pedestrians.
Trails in nature form spontaneously and often in the absence of a leader; they are self-organised dynamic structures
that emerge from the delayed and long-range coupling between the individuals.
The basic mechanism originates in individuals leaving markers on a surface to facilitate the process of foraging~\cite{fraenkel1961orientation, kennedy1986some, blanton2000cellulose, bonner1947evidence, amselem2012stochastic}, mating~\cite{merz2002bacterial, reid2012slime, hoiczyk2000gliding}, or defeating an invader~\cite{lim2015neutrophil}.
Cooperating members of the same species align their motion with and reinforce these markers, often giving rise to the formation of large-scale patterns, for example, a network of trails.
Beyond this qualitative picture, the physics of this kind of trail formation is not well understood and constitutes an intriguing problem of active matter research.

The behaviour of ``active'' (self-propelled) synthetic particles has been subject of intense research in soft matter and statistical physics, with focus on the motion of individual particles \cite{marchetti2013hydrodynamics, kurzthaler2016intermediate} and their response to complex environments
\cite{bechinger2016active, choudhury2017active, mokhtari2019dynamics, chepizhko2013diffusion}.
Concomitantly, strong emphasis was laid on emerging collective patterns, which include
flocking~\cite{toner2005hydrodynamics}, herding~\cite{toner1998flocks}, vortices~\cite{wensink2012meso, grossmann2014vortex, liao2021emergent}, and clustering~\cite{fily2012athermal, redner2013structure, palacci2013living, suma2014motility, schwarz2012phase, mokhtari2017collective}, just to mention a few.
Other than such compact structures, low-dimensional patterns such as isolated trails are relatively underinvestigated.
A notable exception are models for the emergence of pedestrian trail systems connecting certain entry points and destinations \cite{helbing1997modelling, helbing1997active} and traffic phenomena along existing trails \cite{chowdhury2005physics}.

The ability to navigate in gradients of a secondary field, such as a signalling chemical, leads to the phenomenon of chemotaxis, which receives growing interest in physics research \cite{liebchen2018synthetic, meyer2014active, eisenbach2004chemotaxis, mittal2003motility, grima2005strong},
with only few recent works on auto-chemotaxis (e.g., \cite{taktikos2011modeling, amselem2012stochastic, meyer2014active,kranz2016effective}).
Continuum mathematical models of auto-chemotaxis are traditionally based on the Patlak--Keller--Segel equations (PKS), where the local density of agents is coupled to the gradient of the chemical field \cite{keller1970initiation,painter2019mathematical} and which can be derived from a kinetic equation for the distribution of autochemotactic agents \cite{perthame2018fluxlimited}. The PKS model predicts clustering and has been refined to avoid unphysical chemotactic collapse \cite{canizares2017phd}, but a more complex model is required in order to demonstrate trail formation \cite{painter2019mathematical,boissard2013trail}.
We note that many of the experimental studies on chemotaxis as well as the PKS model aim at suspensions of microswimmers or microorganisms, where the surrounding fluid leads to a significant diffusion of the released chemical. This is in contrast to the scale of insects, where pheromone diffusion is comparably small.
Pioneering work on spontaneous trail formation was inspired by the behaviour of ants and relied on cellular automata \cite{edelstein1994simple,edelstein1995trail}, whereas ant models based on agents which adjust their own direction of motion in response to a pheromone have been proposed more recently \cite{boissard2013trail,amorim2019ant}.
From the few available agent-based studies of trail formation, there is no consensus on important questions such as:
What are the essential or minimal requirements for trails to emerge? Which parameters decide whether an area-covering network or a focused trunk trail or no trail at all forms? Does spontaneous trail formation constitute a non-equilibrium phase transition, or is it merely a crossover phenomenon?

In this work, we study trail formation by spontaneous symmetry breaking of a statistically homogeneous state.
To this end, we consider an agent-based model that mimics the behaviour of ants, treating the agents as persistent walkers (self-propelled particles) and the chemicals as pheromone droplets.
Based on extensive computer simulations and a macroscopic dynamic model,
we collect evidences for a putative dynamic phase transition:
An initially disordered distribution of individuals self-organise into a collective state consisting of long-lived trails in the ordered phase (see Supplementary Material for a movie). The study of dynamical phase transitions in systems of self-propelled particles has been a subject of interest since around the turn of the millennium~\cite{vicsek1995novel, toner1995long}. A variety of phenomena including coherent spontaneous large-scale patterns of particles can emerge in such out-of-equilibrium systems. Despite the appealing analogies between some of such observed phase transitions to the transitions in equilibrium, differences are numerous and decisive. In particular, unlike the critical phenomena in thermal equilibrium, for active matter systems there is no trivial knowledge which features of a system are irrelevant details and which are dominant~\cite{binder2021phase}.


\section{Model and methods}
\subsection{Agent--pheromone model}
\label{sec:model}

\begin{figure}
\includegraphics[width=0.3\textwidth]{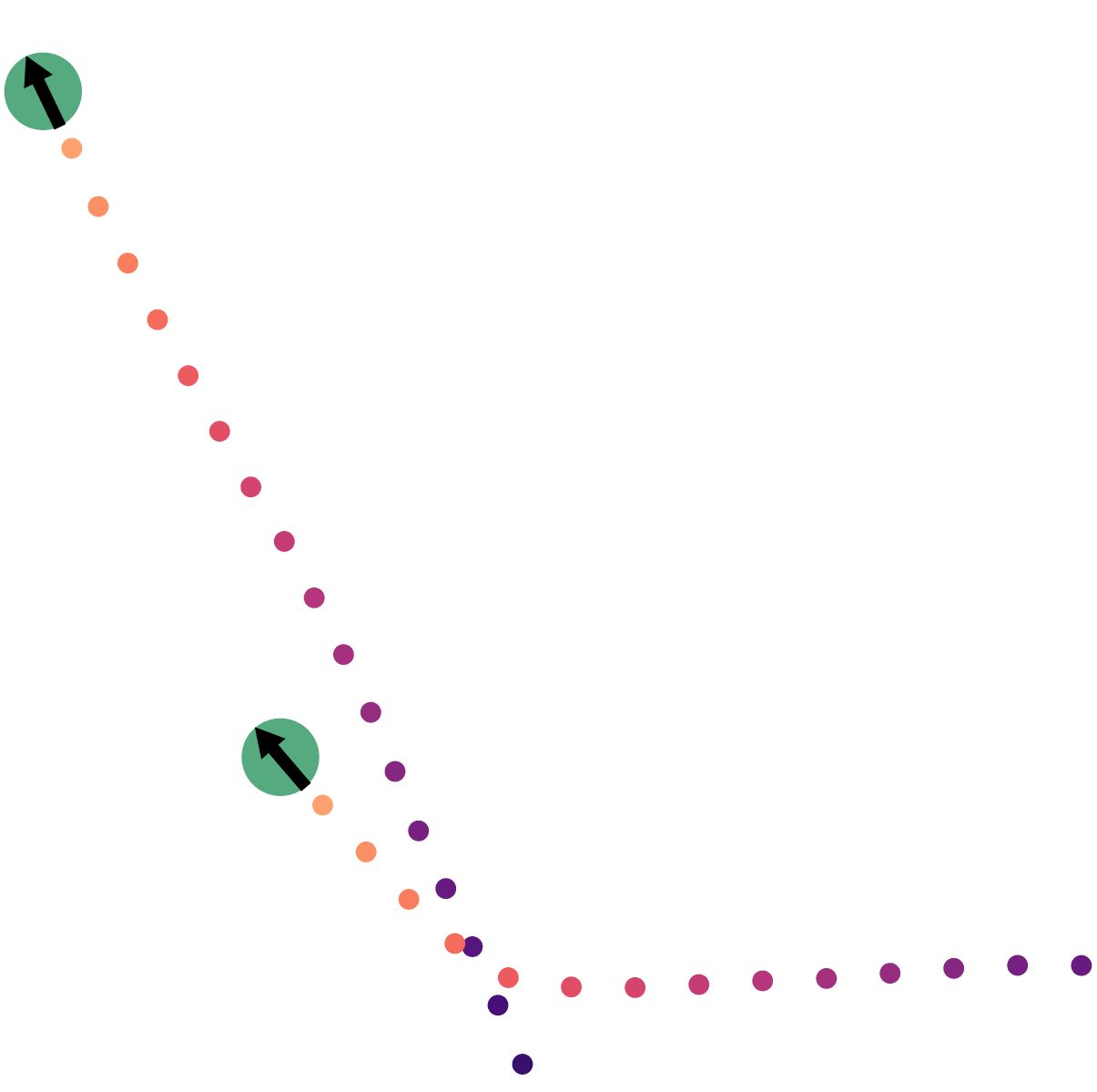}
\caption{Persistent walkers or ``agents'' (green discs) secrete pheromones as they move and experience a torque that aligns their orientation to that of a nearby pheromone trace. The pheromones degrade over time (orange dots turn purple as they age).
}
  \label{fig:model}
\end{figure}

\begin{table}[b]
\begin{tabular}{lcrl}
\toprule
Parameter & Symbol & \multicolumn{2}{c}{Value} \\
\midrule \midrule
disk radius of agent & $R_0$ & \multicolumn{2}{c}{unit of length} \\
rotational diffusion constant & $D_r$ & \multicolumn{2}{c}{unit of time: $\tau_0 := 1/D_r$} \\
\midrule
propulsion speed & $v_0$ & 400 & $R_0\tau_0^{-1}$ \\
pheromone deposition interval & $\tau_d$ & 0.1 & $\tau_0$ \\
pheromone lifetime & $\tau_p$ & 2 & $\tau_0$\\
\midrule
pheromone sensing radius & $r_p$ & 3 & $R_0$ \\
alignment strength & $\mu$ & $(0.1{-}10)$ & $\tau_0^{-1}$ \\
\bottomrule
\end{tabular}
\caption{Parameters of the agent--pheromone model and their values used in the simulations.
The absolute parameter values estimated for the ant \emph{L.~Niger} in Ref.~\cite{boissard2013trail} are recovered approximately by putting $R_0=\SI{0.25}{cm}$ and $v_0 = \SI{2}{cm \per s}$.} 
\label{tab:parameters}
\end{table}

We consider a two-dimensional system composed of $N$ agents that secrete chemicals as they move.
The agents are assumed to walk at constant speed $v_0$ in the direction of random unit vectors, which themselves are subject to a rotational diffusion that is coupled to the pheromones.
The equations of motion for agent $i$ with speed $v_0$ and orientation $\vec n_i=(\cos\phi_i, \sin\phi_i)$ are given by
\begin{equation}
  \dot{\vec r}_i(t) = v_0\vec n_i(t) +\zeta^{-1} \vec F_ i, \qquad
  \dot{\phi_i}(t) = m_i(t) + \eta_i(t) ,
  \label{eq_abm}
\end{equation}
where the $\eta_i(t)$ are independent Gaussian white noises of zero mean and variance $\langle\eta_i(t)\eta_j(t^{\prime})\rangle=2D_r\delta_{ij}\delta(t-t^{\prime})$ in terms of the rotational diffusion constant $D_r$, which serves as unit of time: $\tau_0 := 1/D_r$.
Further, $\zeta$ is the translational friction coefficient, the force $\vec F_i$ on agent $i$ results from a pair repulsion, and $m_i$ is a torque-like contribution due to the alignment interaction with pheromones (see below).
In the first equation, we have neglected translational diffusion relative to the deterministic propulsion, which appears reasonable at the scale of, e.g., insects.
Every agent occupies a disk-shaped area of radius $R_0$ repelling other agents, and this excluded volume effect is realised here through a shifted harmonic force
$\vec F_i = \sum_{j}' K(1-2R_0/r_{ij}) \vec r_{ij}$, where $\vec r_{ij} = \vec r_i - \vec r_j$ and the primed sum runs over neighbouring particles $j\neq i$ with $r_{ij} := |\vec r_{ij}| < 2 R_0$. In the simulations, we used $\zeta^{-1}K = \num{5e5} \tau_0^{-1}$.

We are specifically interested in trail formation in ant colonies. Different species of ants are known to mark the ground with pheromone droplets under different circumstances. Their pheromones often consist of macromolecules that spread by diffusion much slower than the organism moves. To that end, we model the excretions as discrete particles, which are deposited at regular intervals $\tau_d$ and which do not move in space, but evaporate or degrade after a fixed lifetime $\tau_p$ (also see \cref{fig:model}).
A random implementation of the two latter processes, e.g., by independent Poisson processes, appears slightly more realistic, but is unlikely to change the macroscopic behaviour noticeably.

How do the agents interact with the pheromones? The exact form of the response of insects and microorganisms to chemical secretions is not well understood in many cases \cite{perna2012individual}. Yet, there seems to be consensus that the interaction is mediated through a coupling between the walking direction and the trail orientation.
A wide range of ant species reorient upon approaching a trail and move along it towards the direction of freshly deposited chemicals~\cite{aron1993memory, jackson2004trail, czaczkes2015trail}.
For the pheromones in the present model to be able to induce such an effect we assign an intrinsic orientation to the pheromones, which copies that of the agent at the moment of deposition \cite{boissard2013trail}.
The agent--pheromone interaction is implemented as an aligning torque exerted on the agent by the pheromones surrounding it within a disk of radius $r_p$:
\begin{equation}
  m_i = \mu \sum_{j \in J_i} \sin(\theta_j-\phi_i) \,,
  \label{eq_torque}
\end{equation}
where $\theta_j$ refers to the orientation of pheromone~$j$ and $J_i$ is the set of pheromone indices within the sensing radius $r_p$ of agent~$i$.
A positive interaction strength $\mu>0$ causes alignment, whereas $\mu < 0$ would lead to anti-alignment.


The model parameters were chosen to approximately match the values estimated for the black garden ant \emph{(Lasius Niger)} \cite{boissard2013trail, beckers1992trail} and are summarised in \cref{tab:parameters}.
With these parameters, the orientation of the agents is highly persistent relative to their translational motion, expressed by the dimensionless ratio $v_0/(R_0 D_r)=400 \gg 1$.
The pheromone trace of a \emph{single} agent on average is constituted of $\tau_p / \tau_d = 20$ pheromones (\cref{fig:model}), and two successively deposited pheromones are separated by a distance $v_0\tau_d =40 R_0$, which is much larger than the sensing radius $r_p = 3R_0$.
Thus, the average number density of pheromones in the system is $\rho_p = (\tau_p / \tau_d) N L^{-2}$, which multiplied by the sensing area of a single agent yields the mean number of pheromones $N_p = \pi r_p^2 \rho_p$ in the vicinity of an agent (ignoring any interactions). This is an important control parameter, and for $N_p$ varied between 1 and 100 we observe a variety of collective behavior.

\subsection{Simulation details}

For the demanding simulations of the agent-based model, we exploited the massively parallel hardware of high-end graphics processing units (GPUs) \cite{anderson2008general,colberg2011halmd,weigel2012gpu}
and relied on the particle-based simulation package HOOMD-blue~\cite{anderson2008general,glaser2015strong}.
In-house modifications were necessary to include the aligning torque and the deposition and evaporation of pheromones.
The simulation domain was a square of edge length $L=400 R_0$ with periodic boundary conditions applied along both Cartesian directions.
All simulations start from a homogeneous state, i.e., a random distribution of agent positions, independently and uniformly sampled from the domain, and without pheromone droplets.
The equations of motion, \cref{eq_abm,eq_torque}, were solved with the Brownian integrator of the package, which resembles the Euler--Maruyama scheme \cite{Leimkuhler:Molecular}, with integration time step $\delta t=\SI{e-5}{\tau_0}$; the smallness of this number is due to the large ratio $v_0\tau_0/R_0=400$ (see \cref{tab:parameters}). We discarded an initial relaxation phase of $\Delta t = 25\tau_0$ before sampling at intervals of $5\tau_0$ for the stationary averages. For larger systems of size $L=1200R_0$, we discarded a 9 times longer relaxation phase of $\Delta t = \SI{225}{\tau_0}$.
For each set of parameters shown in \cref{fig:phase_diagram}, more than 200 simulations with independent initial configurations were run, consuming more than \num{1400} hours of GPU time in total. Depending on the agent density, the number of agents was varied between $N=160$ and \num{16000} with the number of pheromone particles ranging between \num{3200} and \num{320000}.


\subsection{Identification of trails}
\label{sec:data_analysis}

\begin{figure}
\includegraphics[width=0.6\textwidth]{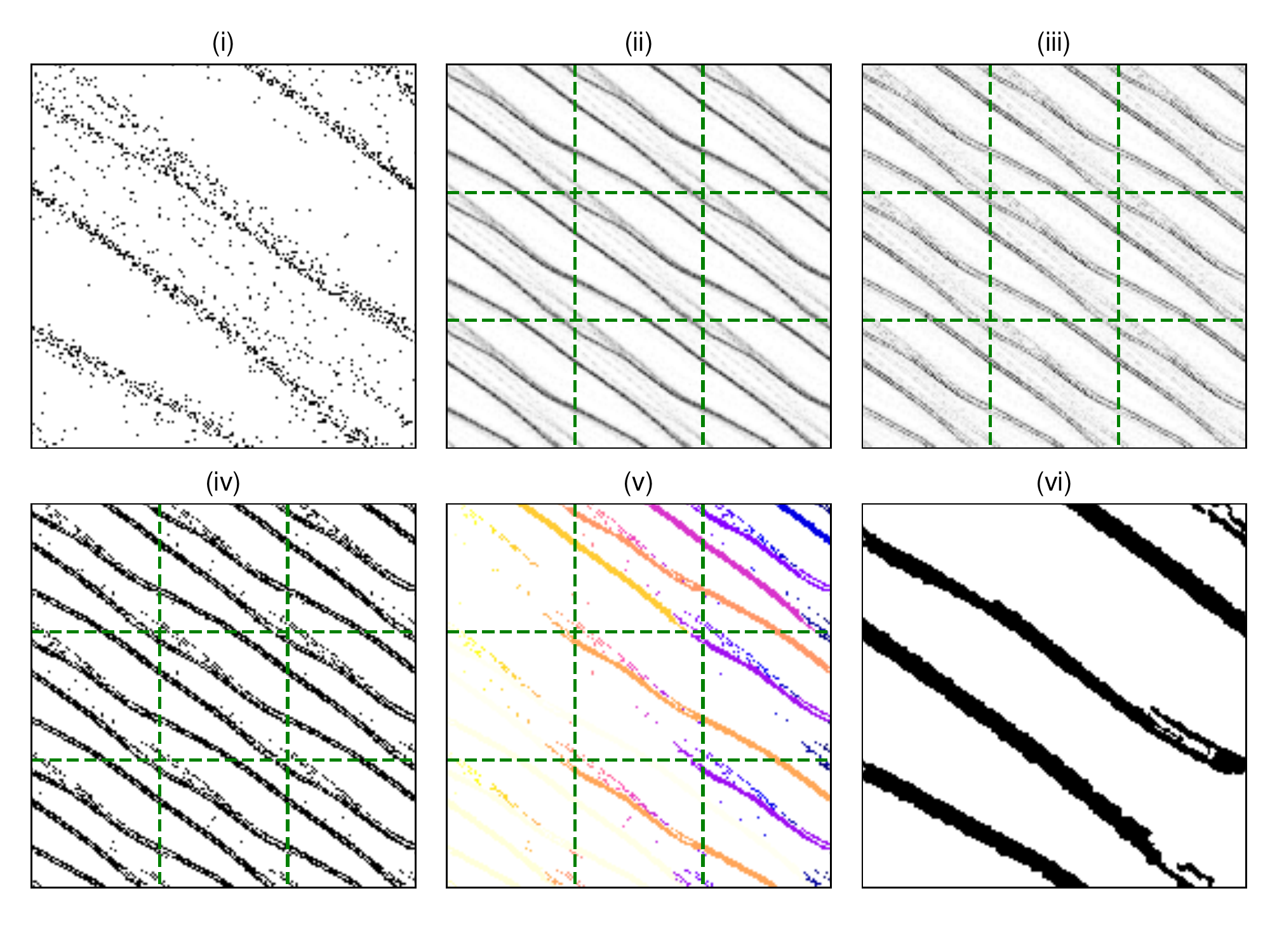}
\caption{Identification of trails by mapping discrete pheromones to a bitmapped field and applying image processing techniques: Gaussian filter, edge detection, thresholding, and segmentation. The trails, defined as the percolating object(s), are coloured in black in the post-processed image (panel (vi)). Each step, indicated by a number on the top of the panel, is explained in \cref{sec:data_analysis}.}
  \label{img_processing}
\end{figure}

We understand ``trails'' as long, thin connected domains filled with pheromone droplets and therefore trail edges will exhibit large gradients in the pheromone concentration (perpendicular to the trail direction).
We developed the following simple image processing procedure to automatically identify trails in simulation outputs for the positions of pheromone particles.
The implementation used the \emph{multi-dimensional image processing package} for Python \cite{van2014scikit}.

\begin{enumerate}
 \item Map the positions of discrete pheromone particles to a discrete pixel field (``bitmap'').
 
 \item Apply a Gaussian kernel. Pixels are squares of size $(L/R_0)^2$, and the blur parameter of the Gaussian filter was taken as 3 pixels.

  \item Edge detection using Sobel filters: Two $3\times 3$ kernels $S_x$ and $S_y$ are convoluted with the original image $A$ to yield approximations $G_x$ and $G_y$ to the intensity gradient along the horizontal and vertical directions respectively.
Periodic boundaries are taken into account by expanding the bitmap and repeating the central bitmap to all 8 neighbouring images.

  \item Convert to a black-and-white image by thresholding of the gradient magnitude $G=\sqrt{G_x^2+G_y^2}$\,;
  a pixel is set to `black' if $G > 0.25 \max(G)$.

  \item Segmentation of the thresholded bitmap into ``trails''. Using a watershed algorithm, objects of connected black pixels are identified and labelled.

  \item Calculate properties of the objects such as their linear extensions, their mass (number of pixels), etc. Identify the trails as the system-spanning or ``percolating'' objects, which we define as those wrapping around the periodic domain, i.e., with an extension along the $x$- or $y$-direction larger than $L/R_0$.
\end{enumerate}

Each step of the procedure is illustrated in \cref{img_processing}.
We do not claim any special status for the details of our image analysis; on the contrary we claim that any reasonable method for identifying trails as 1-dimensional areas of high pheromone concentration surrounded by areas of low concentration would give similar results except possibly in some pathological situations.
In support of this claim we checked that the image processing yields qualitatively similar results if the blur parameter in (ii) and the threshold in (iii) are changed moderately.
The details of the numerical results presented in the remainder of this article certainly will depend on the details of the image processing and we therefore emphasise that our results are qualitative in nature.


\section{Spontaneous trail formation}

\begin{figure*}
\includegraphics[width=.9\linewidth]{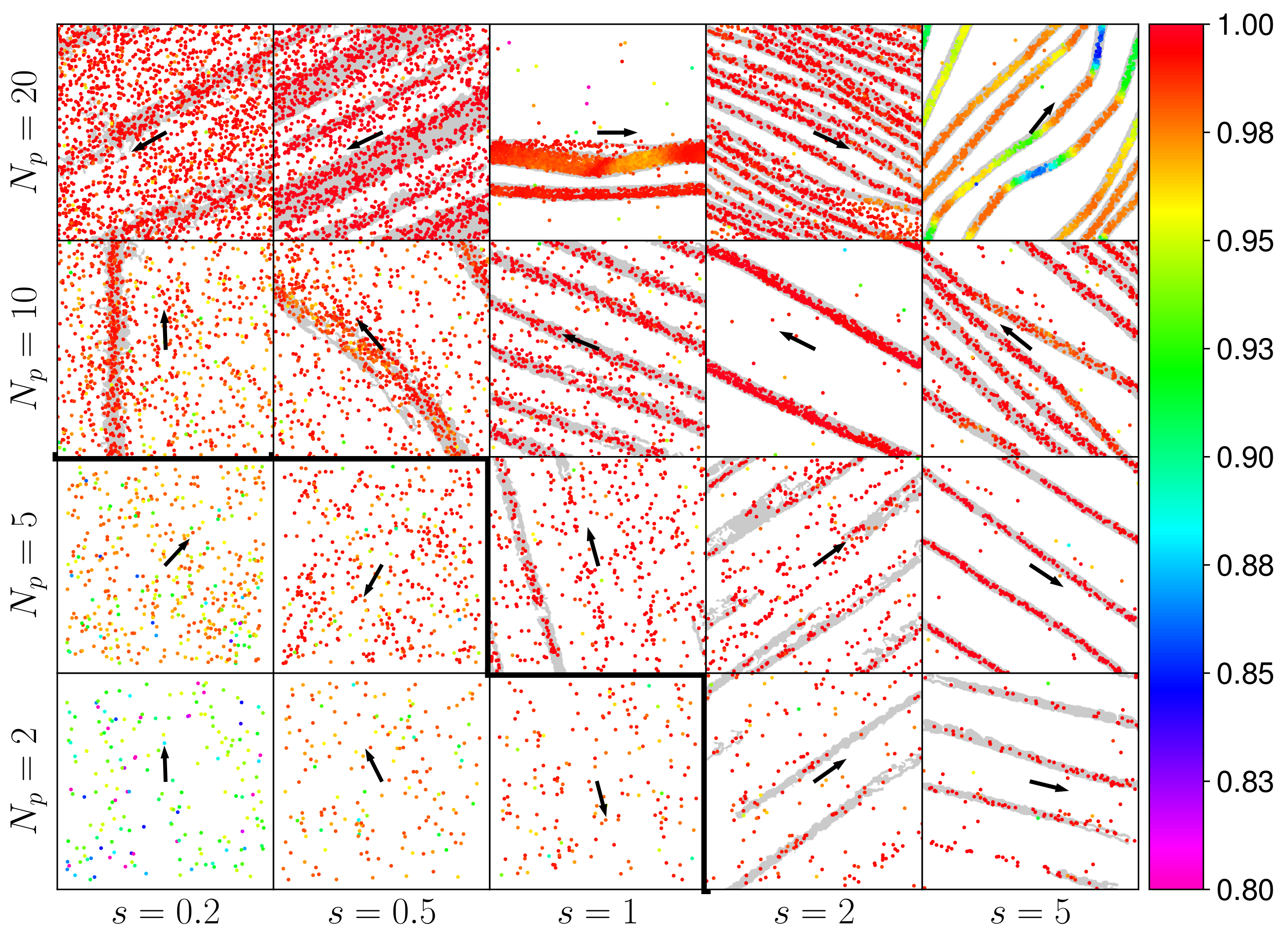}
\caption{Representative configurations of agent positions for different values of $N_p$ and the relative alignment strengths $s=\mu/D_r$,
where $N_p$ is the average number of pheromones interacting simultaneously with an agent (see text);
colours indicate the alignment $\vec n_i\cdot \vec P$ of the walking direction to the global one, $\propto \vec P$ (black arrows), where 1 refers to perfect alignment.
Thick panel borders separate homogeneous, disordered states (bottom left panels) from states with several weak or few trunk trails (top and right panels).
The trails are relatively straight for a large range of parameters, but tortuous in few cases. Light grey stripes represent system-spanning trails that were identified from the distribution of pheromones. 
}
  \label{fig:phase_diagram}
\end{figure*}

We have carried out extensive simulations to obtain a qualitative description of the stationary states that emerge within the investigated model, laying particular emphasis on a putative transition showing spontaneous trail formation.
Relevant parameters are the strength $\mu$ of alignment with the pheromones, the orientational diffusion $D_r$,
and the average number $N_p$ of pheromone droplets within the detection range of an agent.
The first two parameters have antagonistic effects on trail formation and form the dimensionless ratio $s=\mu/D_r$.
With increasing $N_p$, agents are expected to align more effectively to each other, favouring trail formation.
We explored this parameter space for $s\in [0.2,5]$ and $N_p\in [2,20]$ by changing the interaction strength $\mu$ and the number of agents $N$, while keeping all other parameters fixed to their values given in \cref{tab:parameters}.

A chart of typical configurations observed in the stationary regime is shown in \cref{fig:phase_diagram}.
In the dilute case and if interactions are weak (e.g., $s=0.2, N_p = 2$), the spatial distribution of agents remains homogeneous and isotropic in a statistical sense.
For sufficiently large $N_p$ and $s$, agents accumulate along straight lines that wrap around the periodic simulation domain:
persistent and spanning trails develop.
Depending on the control parameters, most of the agents follow these macroscopic trails and thereby reinforce them; on long time spans, the trails are dynamically remodelled and can slowly change their location.
This is very well exposed in \cref{fig:phase_diagram}, for example, at $s=2, N_p = 10$ and at $s=5, N_p = 5$.
For a number of situations, however, we observe that many thin trails emerge that run in parallel instead of merging to a single (or few) trunk trails, even over long times.

\begin{figure}
  \includegraphics[width=\linewidth]{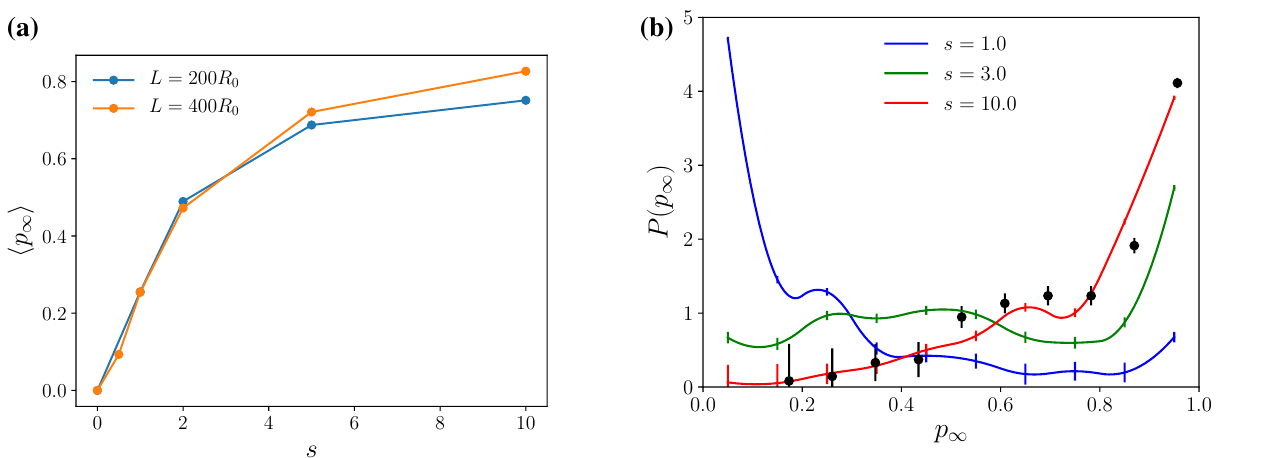}
  \caption{The fraction of pheromones in the spanning trail, $p_{\infty}$, quantifies a putative transition from a homogeneous state to a state with long-lived trails.
  Panel~(a): The ensemble-averaged trail weight $\langle p_{\infty} \rangle$ as function of the relative alignment strength $s=\mu/D_r$ for two different sizes $L$ of the simulation box, as indicated in the legend.
  ~Panel~(b): Probability density of $p_{\infty}$ for three different alignment strengths $s$ and a box size of $L=400R_0$.
  Lines are smooth interpolations through the data points, represented by their error bars.
  Black symbols correspond to results for $s=10$ obtained with a large box of $L=1200R_0$.
  In both panels, $N_p = 5$ and all other parameters were fixed to their values in \cref{tab:parameters}.
  }
  \label{fig:p_inf}
\end{figure}

In order to quantify a possible dynamic phase transition, we introduce the relative weight $p_\infty$ of (the) system-spanning trail(s) as a tentative order parameter. We define it in analogy to the weight of the infinite cluster in percolation theory \cite{benAvraham:DiffusionInFractals,kammerer2008percolation} as
\begin{equation}
	p_{\infty} = \frac{\text{amount of pheromone in spanning trails}}{\text{total amount of pheromones}} \,;
	\label{eq:p_infty}
\end{equation}
within the above detection scheme, $p_\infty$ is calculated as the ratio of the number of pixels in spanning objects and the total number of `black' pixels (see step (iv) of the trail identification algorithm). Its mean value $\langle p_\infty \rangle$
is calculated as a stationary ensemble and time average and it increases monotonically with $s$ for $N_p=5$ fixed, with the fastest variation for $s\approx 0.5{-}1$ (\cref{fig:p_inf}a).
The probability density function of $p_{\infty}$ exhibits a pronounced peak for small and large values of $s$, with the peak and also the mean value shifting from $\langle p_{\infty}\rangle \approx 0$ to $\langle p_{\infty}\rangle \approx 1$ upon increasing $s$ across a certain threshold value (\cref{fig:p_inf}b).
Comparing the results in \cref{fig:p_inf} for different system sizes,
both the variation of $\langle p_\infty \rangle$ with $s$ and the order parameter distribution display only relatively small changes despite the considerable changes of $L$ by factors of 2 and 3, respectively.
This suggests that, for the present parameter set and choice of the order parameter, increasing the system size further would not result in a sharp phase transition.
Yet, we emphasise that increasing the system size to $L=1200 R_0$ does not stop the emergence of long-lived, spanning trails in the stationary state. 

\begin{figure}
\includegraphics[width=.5\textwidth]{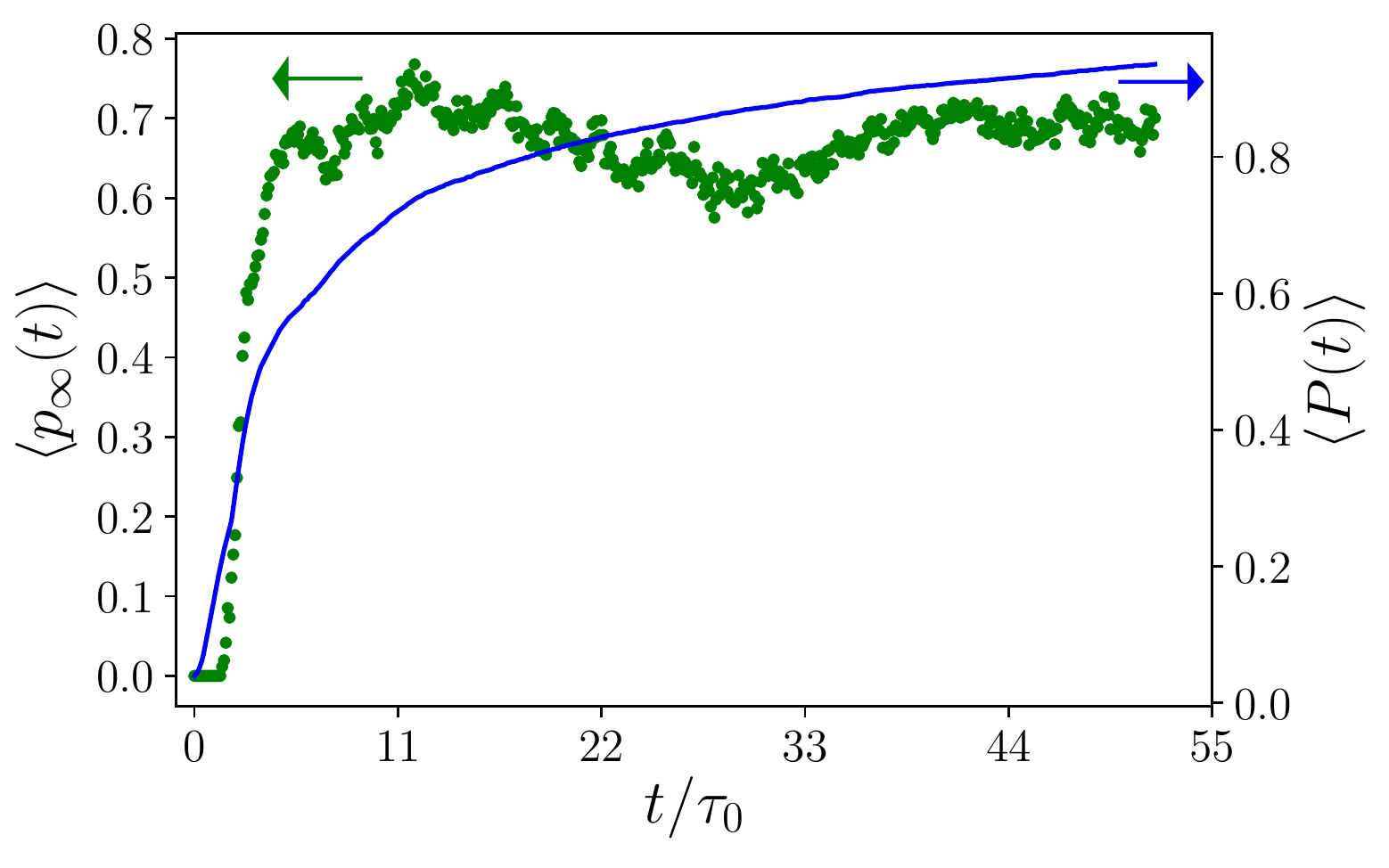}
\caption{Time evolution of the ensemble average of the fraction of pheromones in the spanning trail $p_{\infty}$ (green disks, left axis) and magnitude of the polarisation $\vec P=N^{-1}\sum_{i=1}^N \vec n_i$ (blue, solid line, right axis) for $N_p=5, s=5$.
The average was taken over 100 simulations, which explains the residual statistical fluctuations in $\langle p_\infty(t) \rangle$.
}
  \label{fig:polarity-and-p_infty}
\end{figure}

For all panels shown in \cref{fig:phase_diagram}, the agent system exhibits a well-developed polarisation
$\vec P=N^{-1}\sum_{i=1}^N \vec n_i$ in the steady state, even at low densities and weak alignment interaction.
A comparison of the dynamic evolution of polarisation and that of the spanning trail, upon starting from a homogeneous state, suggests that at moderate values of $N_p$ and $s$, polarisation develops more gradually and more slowly than trails form (\cref{fig:polarity-and-p_infty}).
In particular, polar ordering is not a prerequisite for the formation of trails. Conversely, at low density and interaction strength, in the absence of macroscopic trails, the system is still polar, i.e., shows a coherent movement, which we attribute to the persistent motion of agents with a finite-range alignment interaction.

From the above simulation results, it seems that both trail formation and polarity originate in the aligning interaction between agents and pheromones. To further investigate the importance of the polarity of the pheromones for the emergence of trails, we have run some exemplary simulations for a modification of our model using directed, but apolar pheromones; specifically, we have changed the aligning torque from the expression in \cref{eq_torque} to $m_i=(\mu/2) \sum_{j \in J_i} \sin(2(\theta_j-\phi_i))$. The preliminary results
suggest that macroscopic trails still emerge as in the original model and that the overall behaviour is qualitatively reproduced. We conclude that the polarity of the individual pheromones is not essential for the formation of trails.

\begin{figure}
\includegraphics[width=.5\textwidth]{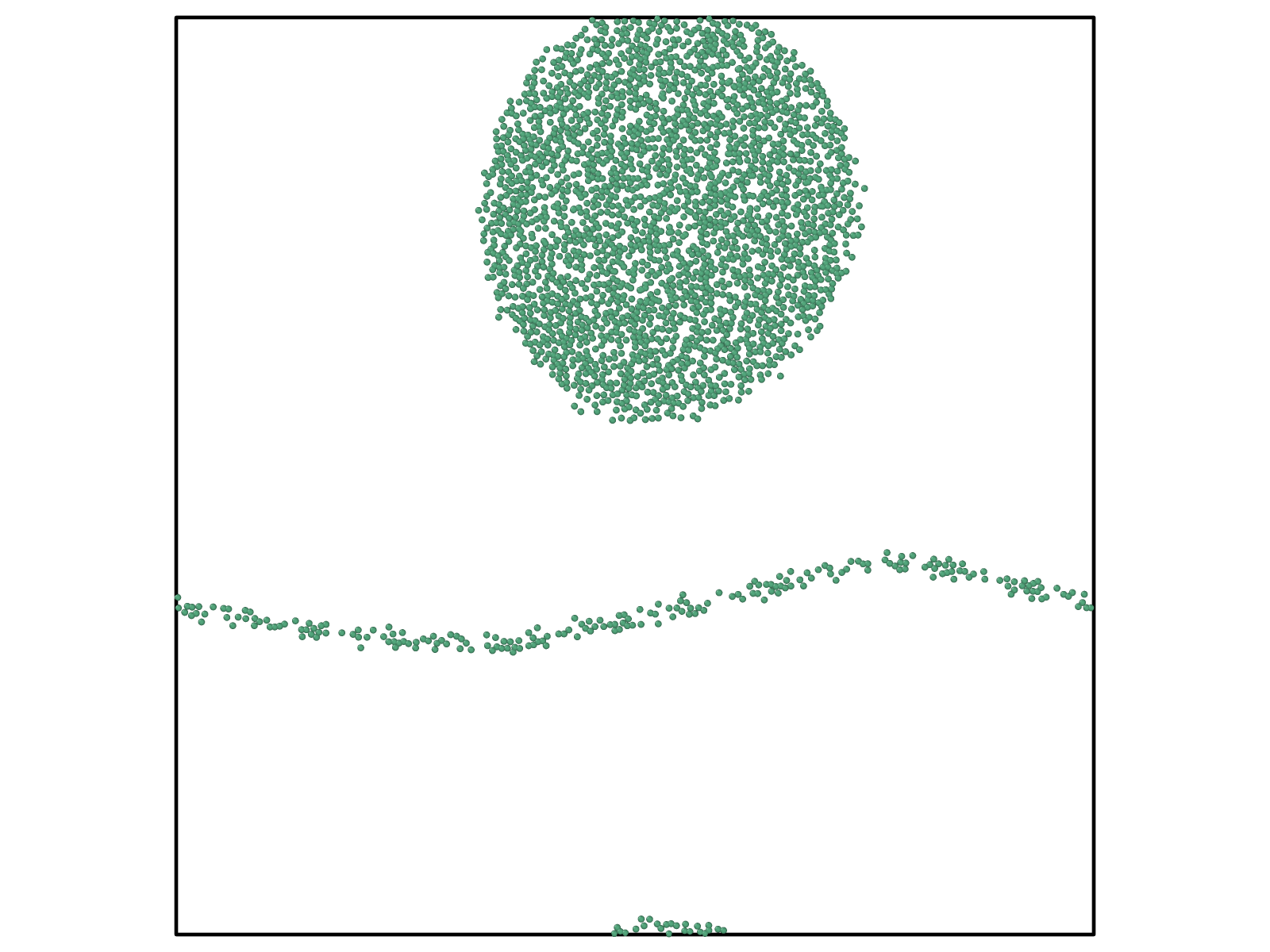}
\caption{In dense systems of strongly aligning agents, large rotating clusters (\emph{ant mills}) develop and can coexist for a long period of time with a spanning trail. The latter is eventually swallowed up by the rotating cluster.
The figure shows a typical configuration of the agent positions (green dots) for $N_p\approx 30$ and $s=10$; the borders indicate the periodic simulation domain with an edge length of $L=400R_0$.
}
  \label{fig:cluster-and-trail}
\end{figure}

A further increase of the alignment strength (for example, at $s=5, N_p = 20$) weakens the polarisation due to the formation of tortuous trails:
While the trails tend to straighten at moderate values of $s$, a large alignment strength $s$ allows the agents to follow kinked and curved trails, to the point where they can follow circular trails. In this case, agents can be trapped in rotating clusters, which grow in size as the agent density is increased (\cref{fig:cluster-and-trail}).
These rotating clusters share some similarities with \emph{ant mills}, a phenomenon in which a group of army ants begin to follow one another, forming a stable and continuously rotating cluster~\cite{beebe1921edge}.
\Cref{fig:cluster-and-trail} exhibits such a large rotating cluster of more than \num{2000} agents, in coexistence with a long-lived spanning trail, which is eventually swallowed up by the mill (see Supplementary Material for a movie).
We conclude that for sufficiently large values of $s$, the trails cease to exist.


\section{Macroscopic dynamic model}
\label{sec:macroscopic_model}

\begin{figure}
\includegraphics[width=\linewidth]{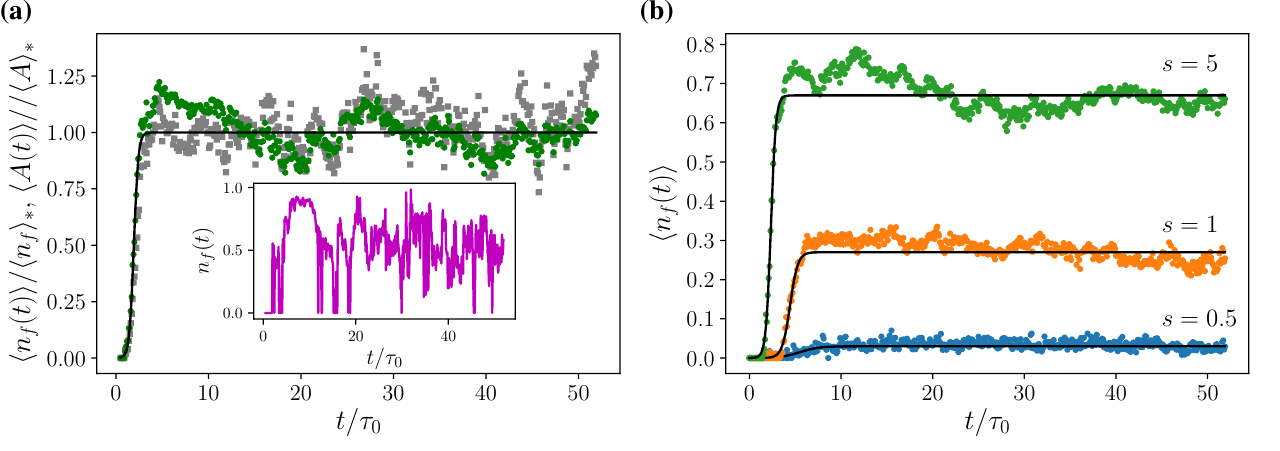}
\caption{Panel~(a): Temporal evolution of the mean fraction of agents following a trail ($\langle n_f(t) \rangle$, green disks) and mean area fraction of trails ($\langle A(t) \rangle$, grey squares), starting from statistically homogeneous and isotropic configurations for parameter values $s=5$ and $N_p=5$. Both quantities have been normalised by their stationary mean values, $\langle n_f \rangle_*$ and $\langle A \rangle_*$, respectively, where $c=\langle A\rangle_*/\langle n_f\rangle_* \approx 0.18$.
The non-equilibrium ensemble averages were taken over $\approx 50$ simulations.
The inset shows $n_f(t)$ as observed in a single simulation.
~Panel~(b): Evolution of the mean fraction of followers for three different relative alignment strengths $s=\mu/D_r$ as indicated,
$N_p=5$ and all other parameters as in \cref{tab:parameters}.
In both panels, solid lines depict the solution to the macroscopic dynamic model [\cref{eq_n_f-decoupled}] using the coefficients given in \cref{tab:fitting}.
}
\label{fig:ave_n_f-vs-t}
\end{figure}

Aiming at an understanding of the dynamics of trail formation, we employ a dynamic model of few collective observables that was proposed as a heuristic description of trail following \cite{edelstein1994simple,edelstein1995trail}.
Ideally, such a macroscopic model can reproduce the evolution of the corresponding ensemble-averaged quantities of the microscopic agent--pheromone model.
To start with, we introduce the sub-population of \textit{followers} as the agents that move along a spanning trail; the number fraction of followers relative to the whole population is denoted by $n_f$.
Within the simulations, the number of followers is calculated as the number of agents which are closer to a trail than the sensing radius $r_p$ (see \cref{sec:data_analysis}).
The rest of the agents, called \emph{searchers}, move in the void space between the trails until they hit a trail, join it, and change their identity to a follower.
The number of followers is subject to change due to the recruitment of searchers to existing trails as well as the dislodging of followers from the trails.
The latter contribution is proportional to the population size of followers, whereas recruitment is proportional to the fraction $1- n_f$ of searchers and to the probability of hitting a trail, which is proportional to the    total area fraction $A$ covered by trails in a simplified approach ignoring spatial information.
The dynamics of the average sub-population size of followers is then described by a gain--loss equation of the form
\begin{equation}
	\frac{d\langle n_f(t) \rangle}{dt} = -\gamma_f \langle n_f(t) \rangle + \alpha \bigl[1-\langle n_f(t) \rangle \bigr] \langle A(t) \rangle,
	\label{eq_n_f}
\end{equation}
where $\langle\cdot\rangle$ denotes the time-dependent ensemble average.
The coefficient $\gamma_f$ is the rate of a follower losing a trail, and $\alpha$ is the rate of recruitment of a searcher to a trail upon hitting it.
The area of trails $A(t)$ decreases as followers leave the trails and the pheromones gradually degrade, and it increases as the followers reinforce the trails.
Its dynamics is tightly coupled to that of $n_f(t)$ as corroborated by the simulation data (\cref{fig:ave_n_f-vs-t}a) and we assumed that
\begin{equation}
  \langle A(t)\rangle = c \,\langle n_f(t)\rangle\,,
  \label{eq_a_prop}
\end{equation}
where the constant of proportionality $c = \langle A\rangle_* / \langle n_f \rangle_*$ is fixed by the stationary mean values of $A$ and $n_f$ (denoted by the subscripts $*$).
We note that the data in \cref{fig:ave_n_f-vs-t}a are averages over merely 50 independent simulations and therefore still display ``mesoscopic'' fluctuations. The latter need to be contrasted from the more rapid changes with partially jump-like behaviour of a single observation (inset of \cref{fig:ave_n_f-vs-t}a).
Substituting $\langle A(t) \rangle$ in \cref{eq_n_f} yields
\begin{equation}
	\frac{d\langle n_f(t) \rangle}{dt} = \tilde{\gamma}\langle n_f(t) \rangle - \tilde{\alpha} \langle n_f(t) \rangle^2 ,
	\label{eq_n_f-decoupled}
\end{equation}
where $\tilde{\gamma} = \alpha c - \gamma_f$ and $\tilde{\alpha} = \alpha c$.
This is a closed non-linear ordinary differential equation in $\langle n_f(t) \rangle$.
A linear equation for the reciprocal, $1/\langle n_f(t) \rangle$, is found by dividing \cref{eq_n_f-decoupled} by $\langle n_f(t) \rangle^2$. Its solution for $t > t_0$ given some initial value $\langle n_f(t_0) \rangle > 0$ at time $t_0$ reads
\begin{equation}
	\langle n_f(t) \rangle = \left\{\frac{e^{-\tilde{\gamma} (t-t_0)}}{\langle n_f(t_0) \rangle} + \frac{\tilde{\alpha}}{\tilde{\gamma}} \left[1-e^{-\tilde{\gamma}(t-t_0)}\right]\right\}^{-1}.
	\label{eq_1_n_f}
\end{equation}

\begin{table}[b]
\tabcolsep=2ex
\begin{tabular}{ccd{3}d{3}}
\toprule
$s$ & $\langle n_f \rangle_*$ &
  \multicolumn{1}{c}{$\tilde{\alpha} \tau_0$} &
  \multicolumn{1}{c}{$\tilde{\gamma} \tau_0$} \\
\midrule \midrule
0.5 & 0.03 & 32.12 & 0.96 \\  
1.0 & 0.27 & 8.76 & 2.37\\  
5.0 & 0.67 & 5.49 & 3.68\\  
\bottomrule
\end{tabular}
\caption{Estimated coefficients of the macroscopic model [\cref{eq_n_f,eq_a_prop}] from fitting the solution \cref{eq_1_n_f}
to the data (\cref{fig:ave_n_f-vs-t}b) for different values of the alignment strengths $s$ and for fixed $N_p=5$, $L=400 R_0$, and all other parameters as in \cref{tab:parameters}.}
\label{tab:fitting}
\end{table}

\Cref{eq_n_f,eq_a_prop} constitute the macroscopic model in the variables $\langle n_f(t) \rangle$ and $\langle A(t) \rangle$, which reduces to \cref{eq_n_f-decoupled}.
To test its applicability as an effective description of the agent--pheromone model, we compare the solution of \cref{eq_1_n_f} to simulation results for $N_p=5$ and $s=5$;
all other parameters were set as in \cref{tab:parameters}.
The macroscopic coefficients $\gamma_f, \alpha$ and $c$ depend in an intricate way on the parameters of the microscopic model and their values cannot easily be inferred from the simulation parameters; a much more detailed derivation of a kinetic equation would be required and this would exceed the scope of the article.
Instead, we estimated these coefficients as follows:
calculating the mean value $\langle n_f \rangle_*$ from the data for $\langle n_f(t) \rangle$ in the stationary regime fixes the ratio $\tilde\gamma/\tilde\alpha = \lim_{t\to\infty} \langle n_f(t) \rangle = \langle n_f \rangle_*$. Then, $\tilde\gamma$ was found by fitting \cref{eq_1_n_f} to the data over the full time range, $t \geq t_0:=0.8 \tau_0$, which also yielded values for $\langle n_f(t_0) \rangle \approx \num{e-4}$.
%
%
The results of the procedure are given in \cref{tab:fitting} for $N_p=5$ fixed and different alignment strengths $s$ and
show very good agreement between the macroscopic model and the simulation data (\cref{fig:ave_n_f-vs-t}b).

The stationary solutions, or fixed points, of \cref{eq_n_f-decoupled} are readily obtained as
$\langle n_f \rangle_* = \langle A \rangle_* = 0$, which means no formation of trails, and
\begin{equation}
	\langle n_f \rangle_* = \frac{\tilde\gamma}{\tilde\alpha} =  1 - \frac{\gamma_f}{c \alpha} \,, \qquad
	\langle A \rangle_* = c \langle n_f \rangle_* \qquad
	\text{if} \quad \gamma_f < c \alpha .
	\label{eq_n_f-stationary}
\end{equation}
The last condition results from the fact that $n_f$ cannot be negative
and provides an approximate criterion for the formation of macroscopic trails.
If one succeeds in connecting the macroscopic model to the microscopic parameters, one would have obtained a rough estimate of the regime where trail formation sets in.
To this end, we ignore a possible renormalisation of the coefficients in the macroscopic limit
and, up to constant prefactors, identify the recruitment and losing rates in \cref{eq_n_f} with the alignment strength and the rotational diffusion coefficient of the agent model, $\alpha \propto \mu$ and $\gamma_f \propto D_r$, respectively.
Similarly, the decay and reinforcement of the trails is connected to the rates of pheromone degradation and deposition, $\tau_p^{-1}$ and $n_f \tau_d^{-1}$, respectively,
which suggests that their ratio determines the area fraction of trails in the stationary state, up to a prefactor:
$\langle A \rangle_* \propto \langle n_f \rangle_* \tau_d^{-1} / \tau_p^{-1}$.
Under this correspondence, the condition in \cref{eq_n_f-stationary} reads
$D_r < C \mu \tau_p / \tau_d$ for some constant $C > 0$.
It suggests that the transition or crossover to the trail-forming regime occurs along a line in the state diagram of the microscopic model that is given by
\begin{equation}
	s N_p \approx \textit{const} \,,
	\label{cond_for_trail_formation}
\end{equation}
using $s=\mu/D_r$ and $N_p \propto \tau_p/\tau_d$ (see \cref{sec:model}).
Indeed, such a reciprocal relationship between $N_p$ and $s$ along the transition line is compatible with the qualitative state diagram in \cref{fig:phase_diagram}.
We emphasise that the above derivation is heuristic and far from rigorous. Nevertheless, it facilitates us with first insight into the conditions under which macroscopic trails can form spontaneously in the agent--pheromone system.

\section{Summary and conclusions}
\label{sec:summary}

We have studied the spontaneous formation of macroscopic trails in populations of persistent walkers that interact via chemical secretions.
Aiming at insight into the generic physical mechanism of this phenomenon, we introduced an agent-based model consisting of walkers that move with constant velocity in a certain direction and deposit pheromone droplets.
The trace of an individual is then marked by several pheromones, each characterized by an orientation along the tangent to the trace.
The walking direction of the agents changes by rotational diffusion and by sensing nearby pheromones due to an alignment interaction with the pheromone orientation.
We stress that the model does not contain any attractive pair potential, neither between pairs of agents nor between agents and pheromones, and the only coupling is to their orientations.
The choice of model parameters was guided by available data on the ant species \emph{L. Niger} \cite{beckers1992trail,boissard2013trail}, yet we note that trail formation in the absence
of preferred destinations such as food sources or the nest is somewhat artificial for this species. Army ants, on the other hand, are known to spontaneously relocate their nest, which involves the formation of a focused trunk trail.

Based on extensive simulations of this model, we have qualitatively characterised the stationary state diagram (\cref{fig:phase_diagram}) that is spanned by the average number $N_p$ of pheromones interacting with an agent and the alignment strength $s$ relative to the rotational diffusion constant.
For all agent densities $\propto N_p$ investigated, we observed clearly that long-lived, macroscopic trails emerge from an initially homogeneous state above a certain, $N_p$-dependent value of the alignment strength $s$;
at lower values of $s$, trails do not persist.
In most cases, the trails are almost straight lines, sometimes present as few, strong trunk trails and often as many thin lines running in parallel. The trails can become tortuous for strong alignment, and at high densities we also observed the occurrence of \emph{ant mills}, i.e., rotating clusters (\cref{fig:cluster-and-trail}).

Inspired by the phenomenon of critical percolation, we examined a large number of agent--pheromone configurations for the existence of a system-spanning trail and considered its weight $p_\infty$ as order parameter of a putative dynamic phase transition from homogeneous to trail-forming phases.
Our data for the ensemble-averaged value $\langle p_\infty \rangle$ as well as the distribution of $p_\infty$ are compatible with such a transition (\cref{fig:p_inf}), yet the signatures of a sharp phase transition are not well developed.

Finally, we considered the early phase of trail formation and studied the growth of the sub-population $n_f$ of trail-following walkers (\cref{fig:ave_n_f-vs-t}).
We employed a simple macroscopic dynamic model for ensemble-averaged observables [\cref{eq_n_f,eq_a_prop}] and demonstrated exemplarily that it can reproduce the evolution of the corresponding quantities of the microscopic model.
From the analysis of the macroscopic model, we inferred an approximate criterion for trail formation [\cref{cond_for_trail_formation}] regarding the effects of agent density and alignment strength, which is in qualitative agreement with the non-equilibrium state diagram (\cref{fig:phase_diagram}).

Summarising, we presented an agent-based model with indirect communication through pheromones that exhibits a rich phenomenology including spontaneous trail formation.
We collected first evidence for a dynamic phase transition and gained insight into the conditions under which macroscopic trails emerge.
The model is not necessarily a minimal one for trail formation as preliminary results for a modification with apolar pheromones (i.e., they have a direction, but no orientation) show qualitatively similar behaviour.
Although a number of open questions remain, it is our hope that this pilot study will stimulate future research to clarify the intriguing physics of this type of active matter.

\section{Acknowledgments}
\label{sec:acknowledgemens}

This research has been funded by Deutsche Forschungsgemeinschaft (DFG, German Research Foundation) under Germany's Excellence Strategy -- MATH+ : The Berlin Mathematics Research Center (EXC-2046/1) -- project ID: 390685689 (subproject EF4-10).

\bibliography{bibfile_literature}

\end{document}